\documentstyle[prl,aps,psfig,multicol]{revtex}
\begin{document}
\draft
\title{Electric field scaling at a $B=0$ metal-insulator transition in two 
dimensions}
\author{S.~V.~Kravchenko, D.~Simonian, and M.~P.~Sarachik}
\address{Physics Department, City College of the City University of New York, 
New York, New York 10031}
\author{Whitney Mason and J.~E.~Furneaux}
\address{Laboratory for Electronic Properties of Materials and 
Department of Physics and Astronomy, University of Oklahoma, 
Norman, Oklahoma 73019}
\date{\today}
\maketitle
\begin{abstract}
The non-linear (electric field-dependent) resistivity of the 2D electron 
system in silicon exhibits scaling as a function of electric field and 
electron density in both the metallic and insulating phases, providing 
further evidence for a true metal-insulator transition in this 2D system at 
$B=0$.  Comparison with the temperature scaling yields separate 
determinations of the correlation length exponent, $\nu\approx1.5$, and the 
dynamical exponent, $z\approx0.8$, close to the theoretical value $z=1$.
\end{abstract}
\begin{multicols}{2}

Conventional wisdom holds that a two-dimensional system of noninteracting 
electrons is always localized in zero magnetic field at zero temperature no 
matter how weak the disorder \cite{aalr}.  If so, there should be no 
metal-insulator phase transition at $B=0$ in an infinite sample in 2D.  
However, recent experiments \cite{kravchenko94,kravchenko95} have shown clear 
signatures of a metal-insulator transition in a high-quality two-dimensional 
electron system (2DES) in silicon in zero magnetic field.  The resistivity, 
$\rho$, was found to scale with temperature, so that
\begin{equation}
\rho\,(T,n_s)=f_1\,(|\delta_n|/T^b)\text{ with } b=1/z\nu.\label{z}
\end{equation}
Here $n_s$ is the electron density, $\delta_n\equiv(n_s-n_c)/n_c$, $n_c$ is 
the critical electron density at the metal-insulator transition, $z$ is the 
dynamical exponent, and $\nu$ is the correlation length exponent.  On the 
insulating side of the transition, the resistivity was found to have a 
temperature dependence characteristic of hopping in the presence of a Coulomb 
gap in the density of states \cite{coulomb} due to strong electron-electron 
interactions.  Such scaling behavior is characteristic of a true phase 
transition and implies that there is a diverging correlation length, 
$\xi\sim|\delta_n|^{-\nu}$, at the transition.  The behavior strongly 
resembles the superconductor-insulator phase transition in thin disordered 
films \cite{liu91,kapitulnik95}, as well as the phase transition between 
quantum Hall liquid and insulator \cite{wang94,shahar95,sondhi,wong95}.  These 
results are in apparent contradiction with the scaling theory \cite{aalr} 
which, however, ignores electron-electron interactions.  In the 2DES in 
silicon, electron-electron interactions provide the dominant energy, $E_{ee}$, 
at low electron densities:  at $n_s\sim10^{11}$~cm$^{-2}$, $E_{ee}\sim5$~meV 
while the Fermi energy is only $\sim0.6$~meV.  For comparison, in GaAs/AlGaAs 
heterostructures these energies are approximately equal to each other at the 
same electron density:  $E_{ee}\approx E_F\approx3.5$~meV.  We note that 
recent theoretical studies \cite{shep,efros} have shown that strong 
electron-electron interactions can cause delocalization.

The resistivities reported in Refs.~\cite{kravchenko94,kravchenko95} were 
obtained in the linear regime, {\em i.e.}, in the limit of zero electric 
field, $E\rightarrow0$.  When the electric field is strong, however, the 
effective temperature of the electrons becomes different from the 
lattice temperature.  A general scaling analysis of the Coulomb interacting 
quantum critical point \cite{girvin} shows that the resistivity should also 
scale with the electric field, but the scaling exponent in this case will be 
$1/[(z+1)\nu]$ (instead of $1/z\nu$ in the case of temperature scaling) so 
that
\begin{equation}
\rho\,(E,n_s)=f_2\,(|\delta_n|/E^a)\text{ with } a=1/[(z+1)\nu].\label{z+1}
\end{equation}
Knowing both $z\nu$ and $(z+1)\nu$, one should be able to determine the 
exponents $z$ and $\nu$ separately.  It is worth noting that this was done 
recently for the superconductor-insulator transition in thin disordered 
films \cite{kapitulnik95}.

In this Letter we report measurements of the nonlinear resistivity of a 
two-dimensional electron system as a function of electric field at $B=0$.  
Scaling is observed as a function of electric field and electron density 
about a critical point which corresponds to the same critical density about 
which {\em temperature} scaling was reported earlier 
\cite{kravchenko94,kravchenko95}.  By comparing the electric field scaling 
with the temperature scaling, separate determinations (rather than their 
product) were obtained for the correlation length exponent, $\nu\approx1.5$, 
and the dynamical exponent, $z\approx0.8$, close to the theoretical value 
$z=1$ for strongly interacting Coulomb systems.  At the transition 
($n_s=n_c$), the resistivity is independent of electric field and close to 
$3h/e^2$.  These data provide additional strong evidence for a true 
metal-insulator transition in high-mobility 2DES in silicon in the absence 
of a magnetic field.

The samples used in this work were high-mobility silicon 
metal-oxide-semiconductor field-effect transistors (MOSFET's) with maximum 
electron mobility $\mu^{max}\approx35-40,000$~cm$^2$/Vs similar to those used 
in Refs.~\cite{kravchenko94,kravchenko95}.  The resistivity was determined as 
a function of electric field by measuring current versus voltage ($I-V$ 
curves)
\vbox{
\vspace{0.15in}
\hbox{
\hspace{-0.20in}
\psfig{file=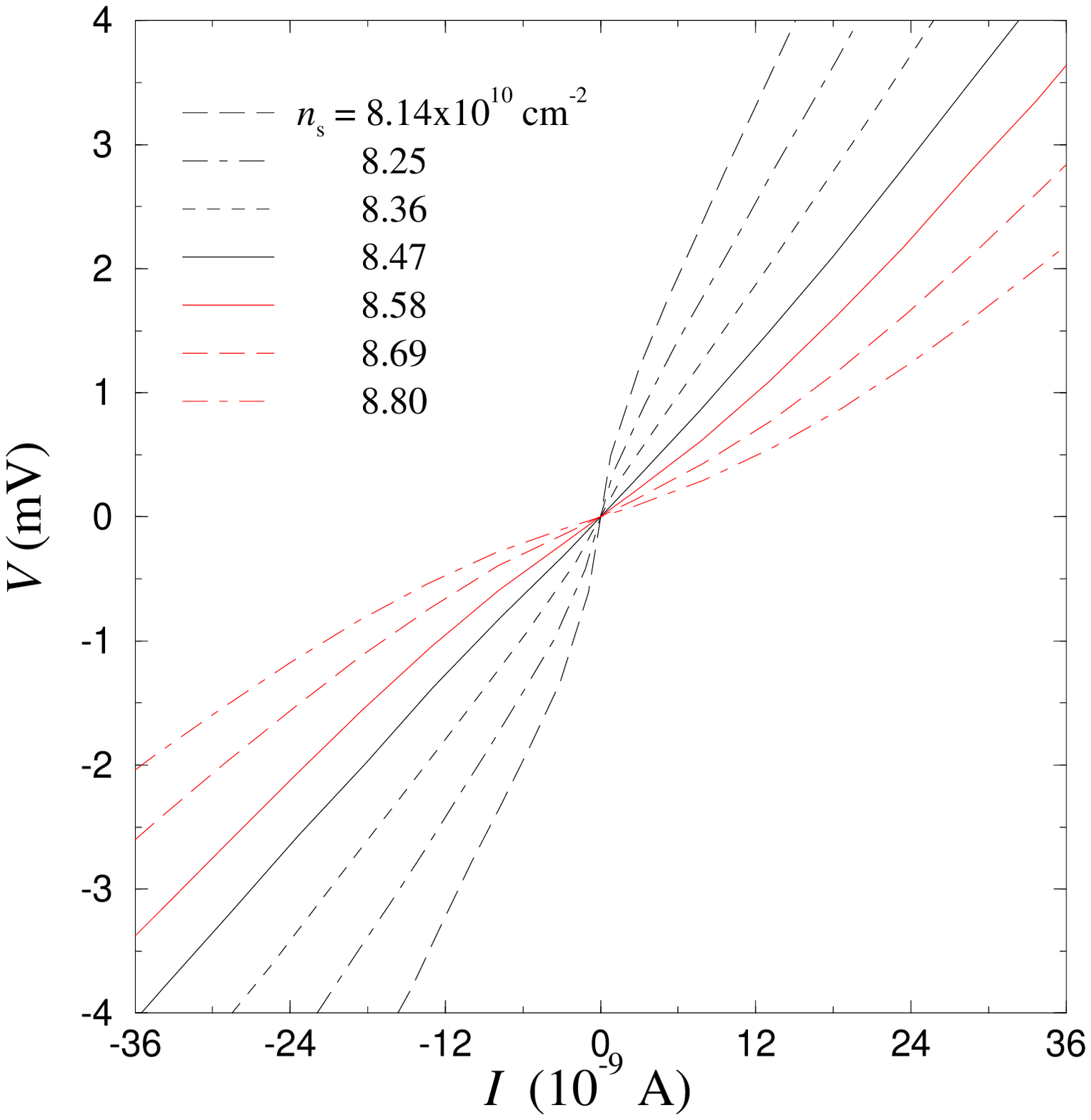,width=3.5in,bbllx=1.5in,bblly=1in,bburx=7.75in,bbury=9.25in,angle=-90}
}
\vspace{0.15in}
\hbox{
\hspace{-0.15in}
\refstepcounter{figure}
\parbox[b]{3.4in}{\baselineskip=12pt \egtrm FIG.~\thefigure.
Nonlinear current-voltage curves at several electron densities around the 
metal-insulator transition.  $T=0.22$~K.
\vspace{0.10in}
}
\label{1}
}
}
for many values of $n_s$ using a standard four-terminal dc technique.  
Typical curves are shown 
in Figure~\ref{1}.  The middle, solid curve is close to linear and separates 
all $I-V$ curves into two groups exhibiting different types of nonlinear 
behavior.  Note that there is a remarkable symmetry about the middle curve.  
This symmetry is reminiscent of that observed for the $I-V$ curves in the 
vicinity of the QHE-insulator phase transition \cite{shahar95,sondhi} and is 
consistent with the existence of a critical point at a critical electron 
density $n_c$.  In Ref.~\cite{sondhi}, the symmetry was attributed to 
charge-flux duality.  In our case, its physical origin is unclear and merits 
theoretical study.

Curves of resistivity $\rho$ {\it vs} electric field $E$ for different 
electron densities are shown in Fig.~\ref{2} (not all curves measured are 
shown in order to avoid too high a density of points).  Here, the resistivity 
is determined from $\rho=(V/I)\cdot(W/L)$ ($W$ is the sample width and $L$ is 
the distance between potential contacts) and the electric field $E=V/L$.  
Again, there is clearly a critical electron density which separates two 
distinct density regions characterized by different types of resistivity 
behavior as a function of electric field.  At the critical point, the 
resistivity (solid curve in Fig.~\ref{1}, closed circles in Fig.~\ref{2}) is 
virtually independent of the electric field and close to $3h/e^2$.  All curves 
below this line are characterized by $d\rho/dE>0$ while all curves above are 
characterized by $d\rho/dE<0$.  The observed critical behavior of $\rho$ as a 
function of $E$ is very similar to the critical behavior of the resistivity 
with {\em temperature} observed near the metal-insulator transition in 2DES in 
silicon \cite{kravchenko94,kravchenko95} (with the only difference that the 
resistivity at the critical point is $E$-independent but {\em not} 
$T$-independent at $T\gtrsim2$~K).  It also resembles the behavior found near 
the superconductor-insulator phase transition in thin metallic films 
\cite{liu91} and the quantum Hall effect-insulator (QHE-I) transition in 
GaAs/AlGaAs heterostructures \cite{wang94,shahar95,wong95}.
\vbox{
\hbox{
\psfig{file=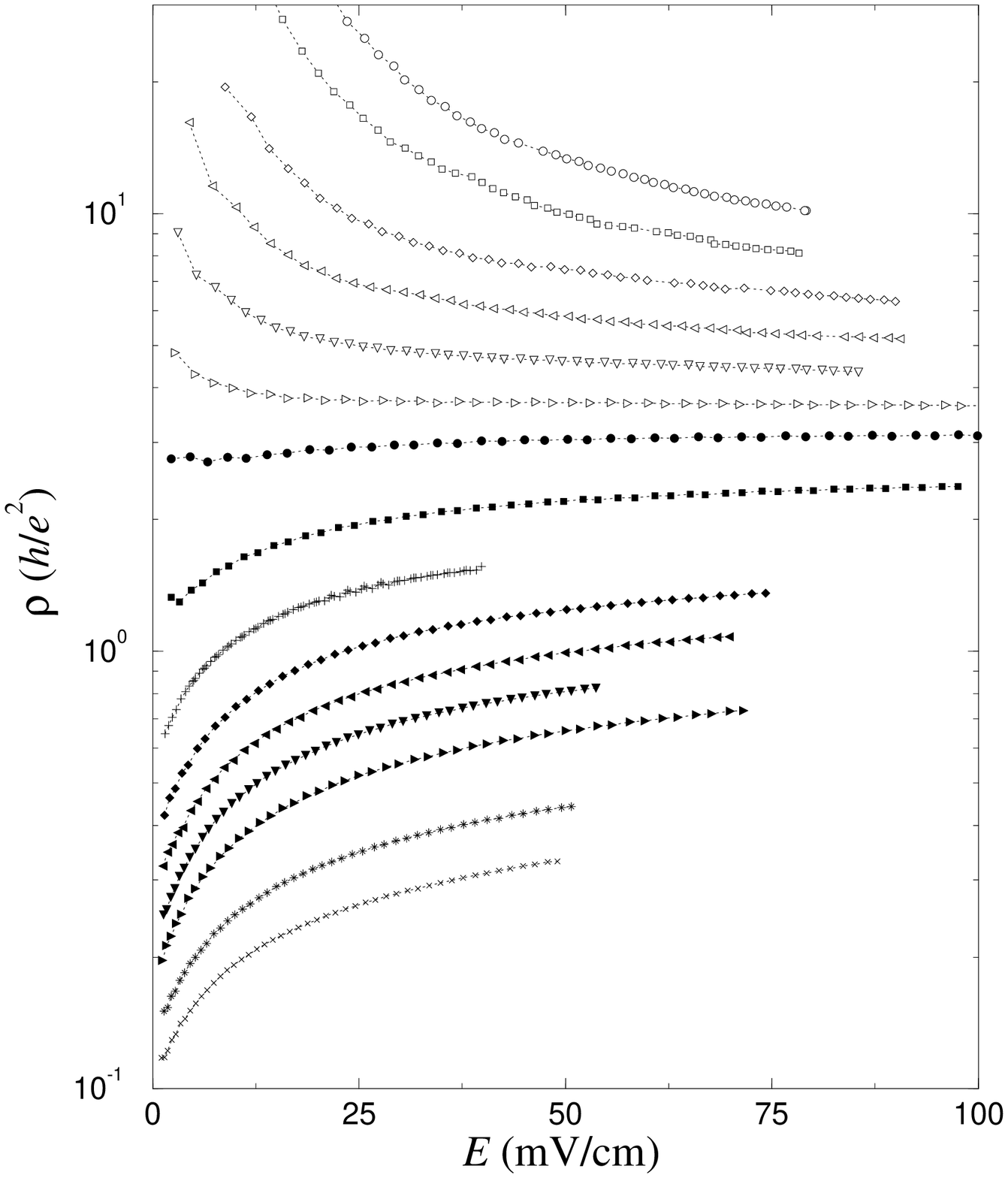,width=3.3in,bbllx=.5in,bblly=1.25in,bburx=7.25in,bbury=9.5in}
}
\vspace{0.15in}
\hbox{
\hspace{-0.15in}
\refstepcounter{figure}
\parbox[b]{3.4in}{\baselineskip=12pt \egtrm FIG.~\thefigure.
Resistivity as a function of electric field at $n_s=7.81$, $7.92$, 
$8.03$, $8.14$, $8.25$, $8.36$, $8.47$, $8.70$, $8.91$, $9.13$, $9.35$, 
$9.57$, $9.79$, $10.34$, and $10.78\times10^{10}$~cm$^{-2}$ at $T=0.22$~K.
\vspace{0.10in}
}
\label{2}
}
}

A remarkable property is that plotting the resistivity against the scaling 
variable, $|\delta_n|/E^a$, causes all the curves to collapse onto two 
distinct branches, as shown in Fig.~\ref{3}~(a).  Note that points for 
$E\rightarrow0$, where $\rho$ saturates due to finite temperature, have been 
omitted.  As expected, the scaling fails at electron densities far from the 
critical point where the system is no longer in the critical regime, as seen 
on the lower curve in Fig.~\ref{3}~(a).  The upper branch corresponds to 
$n_s<n_c$ and the lower one to $n_s>n_c$.  At the transition ($n_s=n_c$), the 
resistivity is close to $3h/e^2$.  It is interesting to note that almost the 
same value was recently reported in Ref.~\cite{valles} for a transition 
between weak and strong localization in disordered metallic films.  The 
exponent $a$ was varied to obtain the best visual collapse, yielding 
$a=0.37\pm0.01$.

Repeating the procedure reported in Refs.~\cite{kravchenko94,kravchenko95}, the 
temperature dependence of the resistivity measured in the linear regime 
($E\rightarrow0$) was used to obtain scaled curves of the resistivity as a 
function of the {\em temperature}-dependent scaling variable, 
$|\delta_n|/T^b$.  This is shown in Fig.~\ref{3}~(b), where it is evident that 
all curves again collapse onto two distinct branches, confirming the earlier 
results.  We note that the data scale well only at temperatures below 1~K, 
presumably because at higher temperatures the system is outside the critical 
regime.  The best collapse was achieved for $b=0.83\pm0.08$.

As mentioned earlier, these two scaling analyses allow
\vbox{
\hbox{
\hspace{0.10in}
\psfig{file=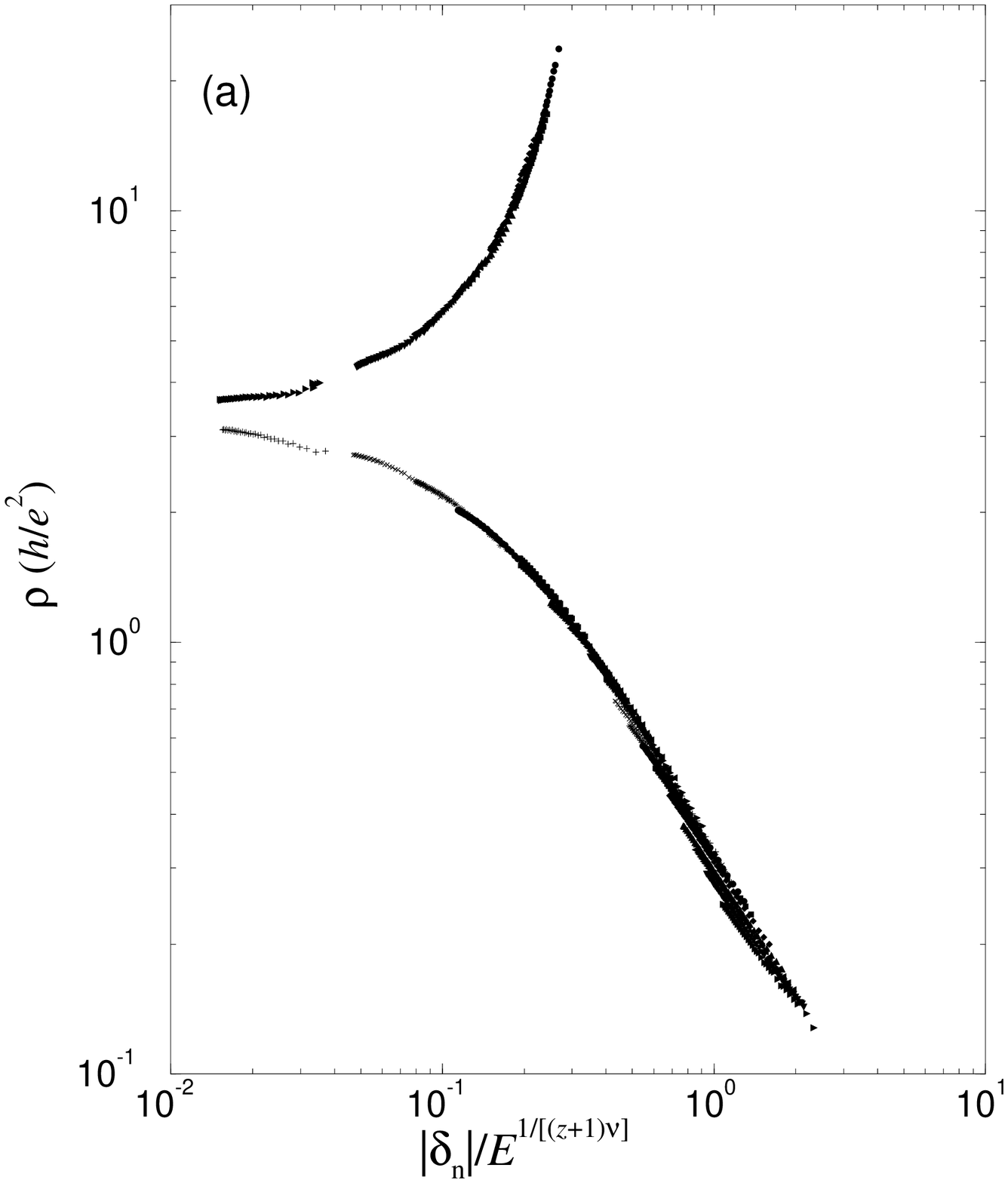,width=3.12in,bbllx=.5in,bblly=1.25in,bburx=7.25in,bbury=9.5in}
}
\vspace{0.15in}
\hbox{
\hspace{-0.15in}
}
}
\vbox{
\vspace{0.1in}
\hbox{
\hspace{0.10in}
\psfig{file=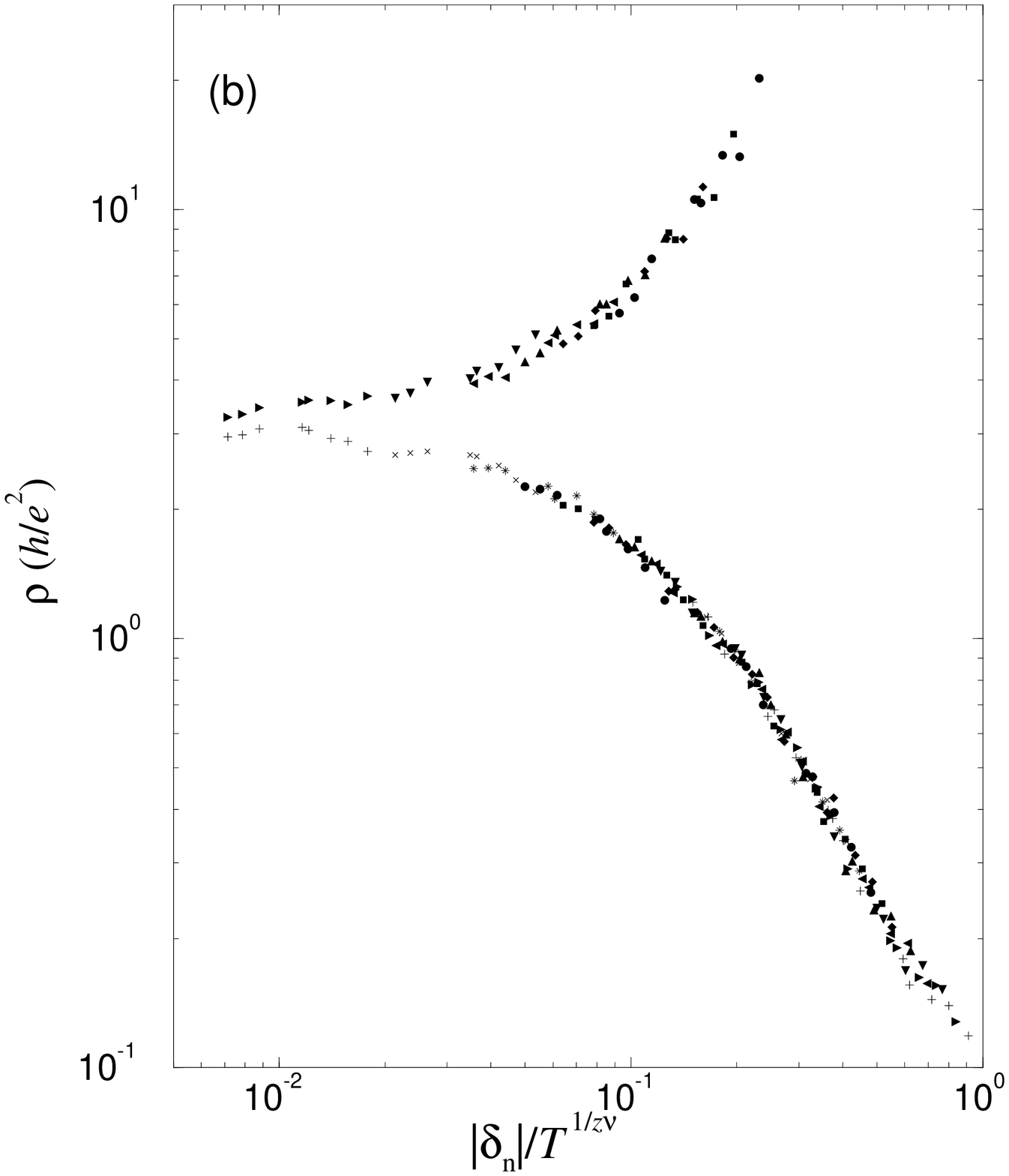,width=3.12in,bbllx=.5in,bblly=1.25in,bburx=7.25in,bbury=9.5in}
}
\vspace{0.15in}
\hbox{
\hspace{-0.15in}
\refstepcounter{figure}
\parbox[b]{3.4in}{\baselineskip=12pt \egtrm FIG.~\thefigure.
(a)~Demonstrating scaling with electric field, the resistivity at 0.22~K is 
plotted as a function of $|\delta_n|/E^a$ for $a=1/[(z+1)\nu]=0.37$.
(b)~Demonstrating scaling with temperature, the linear resistivity 
($E\rightarrow0$) is shown as a function of $|\delta_n|/T^b$ for 
$b=1/z\nu=0.83$.  Electron densities are in the range $7.81\text{ to 
}10.78\times10^{10}$~cm$^{-2}$.
\vspace{0.10in}
}
\label{3}
}
}
one to determine both 
the dynamical exponent and the correlation length exponent.  The values 
obtained in our experiments are $\nu=1.5\pm0.1$ and $z=0.8\pm0.1$.  
While the correlation length exponent has not been theoretically predicted 
for this system, the value of the dynamical exponent in a strongly interacting 
2D system was predicted to be $z=1$~\cite{girvin}, close to what we observe.  
In the majority of interacting 2D systems where it has been determined, the 
exponent $z$ has also been found to be close to 1 (see, {\it e.g.}, 
Ref.~\cite{kapitulnik95} for the case of a superconductor-insulator transition 
and Ref.~\cite{wei94} for the transition between two neighboring QHE 
plateaux).  The same result for the case of hopping conductance follows from 
theory of Polyakov and Shklovskii \cite{polyakov} who found the ratio of
scaling exponents $a/b=1/2$ (see Eqs.~\ref{z} and \ref{z+1}) which is 
equivalent to $z=1$.

We note that the above considerations were based on the assumption that there 
is a diverging correlation time at the critical point which controls the 
dynamics \cite{girvin}.   An alternative picture involving hydrodynamic heating 
was developed in Ref.~\cite{girvin2} which assumes that the electron 
temperature is governed by phonon emission.  In the case of silicon, 
this model gives $a/b=1/3$ rather than 
$a/b=1/2$. Some influence of phonons may be responsible for the ratio $a/b$ 
somewhat less than 1/2 ($a/b\approx0.45$) found in the present work.  A 
crossover between the two regimes will be reported elsewhere 
\cite{unpublished2}.

In summary, we have shown that in the absence of a magnetic field, the 
resistivity of the 2DES in high-quality silicon MOSFETS scales as a function 
of electric field and electron density, exhibiting critical behavior about 
the same point as the temperature scaling.  At the critical point the 
resistivity is independent of electric field and close to $3h/e^2$.  
Comparison of scaling of the resistivity with temperature and scaling with 
electric field yields separate determinations of the correlation length 
exponent, $\nu\approx1.5$, and the dynamical exponent, $z\approx0.8$, close 
to the theoretical value $z=1$.   These results provide additional strong 
evidence of a true metal-insulator phase transition in a high-quality 
2DES in silicon in zero magnetic field.

We would like to thank S.~M.~Girvin, V.~M.~Pudalov, T.~V.~Ramakrishnan, 
and B.~I.~Shklovskii for helpful discussions, and Lucianne Walkowicz for help 
with data handling and preparation of the manuscript.  This work was supported 
by the US Department of Energy under Grant No.\ DE-FG02-84ER45153. 
\end{multicols}
\end{document}